\documentclass[fleqn, usenatbib]{mnras}

\usepackage{newtxtext,newtxmath}

\usepackage[T1]{fontenc}

\DeclareRobustCommand{\VAN}[3]{#2}
\let\VANthebibliography\thebibliography
\def\thebibliography{\DeclareRobustCommand{\VAN}[3]{##3}\VANthebibliography}

\usepackage{graphicx}	
\usepackage{amsmath}	

\usepackage{listings}
\usepackage{color}
\definecolor{dkgreen}{rgb}{0,0.6,0}
\definecolor{gray}{rgb}{0.5,0.5,0.5}
\definecolor{mauve}{rgb}{0.58,0,0.82}
\definecolor{golden}{rgb}{0.86,0.65,0.01}
\title[High mass functions in Gaia DR3]{What are the spectroscopic binaries with high mass functions near the {\it Gaia} DR3 main sequence? }

\author[El-Badry \& Rix]{
Kareem El-Badry$^{1,2,3}$\thanks{E-mail: kareem.el-badry@cfa.harvard.edu}  
and Hans-Walter Rix$^3$\\
$^{1}$Center for Astrophysics $|$ Harvard \& Smithsonian, 60 Garden Street, Cambridge, MA 02138, USA\\
$^{2}$Harvard Society of Fellows, 78 Mount Auburn Street, Cambridge, MA 02138\\
$^{3}$Max-Planck Institute for Astronomy, K\"onigstuhl 17, D-69117 Heidelberg, Germany\\
}

\date{\vspace{-1.0cm}}

\pubyear{2022}

\begin{document}
\label{firstpage}
\pagerange{\pageref{firstpage}--\pageref{lastpage}}
\maketitle

\begin{abstract}
The 3rd data release of the {\it Gaia} mission includes orbital solutions for $> 10^5$ single-lined spectroscopic binaries, representing more than an order of magnitude increase in sample size over all previous studies. This dataset is a treasure trove for searches for quiescent black hole + normal star binaries. We investigate one population of black hole candidate binaries highlighted in the data release: sources near the main sequence in the color-magnitude diagram (CMD) with dynamically-inferred companion masses $M_2$ larger than the CMD-inferred mass of the luminous star. We model light curves, spectral energy distributions, and archival spectra of the 14 such objects in DR3 with high-significance orbital solutions and inferred $M_2 > 3\,M_{\odot}$.  We find that 100\% of these sources are mass-transfer binaries containing a highly stripped lower giant donor ($0.2 \lesssim M/M_{\odot} \lesssim 0.4$) and much more massive ($2 \lesssim M/M_{\odot} \lesssim 2.5$) main-sequence accretor. The {\it Gaia} orbital solutions are for the donors, which contribute about half the light in the {\it Gaia} RVS bandpass but only $\lesssim 20\%$ in the $g-$band. The accretors' broad spectral features likely prevented the sources from being classified as double-lined.
The donors are all close to Roche lobe-filling ($R/R_{\rm Roche\,lobe}>0.8$), but modeling suggests that a majority are detached  ($R/R_{\rm Roche\,lobe}<1$). Binary evolution models predict that these systems will soon become detached helium white dwarf + main sequence ``EL CVn'' binaries. Our investigation highlights both the power of {\it Gaia} data for selecting interesting sub-populations of binaries and the ways in which binary evolution can bamboozle standard CMD-based stellar mass estimates. 
\end{abstract}

\begin{keywords}
binaries: spectroscopic -- binaries: close -- binaries: eclipsing 
\vspace{-0.5cm}
\end{keywords}



\section{Introduction}
\label{sec:intro}


Wide-field spectroscopic surveys have begun to mine the Milky Way for binary stars {\it en masse}. Characterization of spectroscopic binaries is a resource-intensive process because about a dozen individual epochs of spectroscopy are typically required to reliably constrain orbits. Most surveys obtain only one or a few epochs for most targets. Thus, although surveys have {\it identified} many binaries via radial velocity (RV) variability and double lines \citep[e.g.][]{Matijevic2011, Merle2017, Elbadry_2018, Qian2019, Traven2020, Kounkel2021}, the number of well-constrained binary {\it orbits} from wide field surveys was until very recently only of order a thousand \citep[e.g.][]{Price-Whelan2020}. The bulk of all published spectroscopic binary orbits (about $10^4$ in total) instead came from many heterogeneous studies of individual objects and small curated samples  (see \citealt{Pourbaix2004}, for a review). Many of these studies targeted objects recognized as being ``interesting'' for various reasons (e.g., light curve variability, emission lines, chemical peculiarity, UV or X-ray excess, cluster membership, etc.), leading to a rather ill-defined selection function. 

The 3rd {\it Gaia} data release \citep[][]{Valenari2022} dramatically changes the observational landscape of binary star science. Among other data products, it includes 186,905 orbital solutions for spectroscopic binaries, representing more than an order of magnitude increase in sample size over all previous studies. These solutions are obtained by fitting multi-epoch spectra from the Radial Velocity Spectrometer (RVS) instrument \citep[][]{Cropper2018, Katz2022}, which covers the wavelength range of 846-870 nm with spectral resolution $R\approx 11,500$ and $\approx20$ epochs for typical binaries. Both single-lined (``SB1'') and double-lined (``SB2'') solutions are included in DR3 \citep[][]{Arenou2022}. 

The unprecedented volume of the {\it Gaia} spectroscopic binary sample enables searches for classes of binaries that were rare or absent in previous datasets. One such class, which has been subject of intense investigation in recent years, is binaries containing quiescent black holes. Searches for these systems have combed the Galaxy for decades \citep[e.g.][]{Guseinov1966, Trimble1969} and have gained steam recently thanks in part to interest in the progentitors of gravitational wave sources. But to date, the only apparently unambiguous discoveries have been in globular clusters \citep[][]{Giesers2019}.

Here, we explore one sub-population of black hole candidate binaries that can be selected from {\it Gaia} DR3: single-lined binaries near the main sequence in the CMD with implied companion masses that (a) are larger than the CMD-implied mass of the source and (b) are larger than 3\,$M_{\odot}$. \citet{Arenou2022} identified a population of 11 such sources as potentially promising for containing dormant black holes (their Table 10) but were careful to note clearly that other explanations are possible. This paper carries out a more detailed analysis of these sources. We conclude that they are Algol-type binaries near the end of the mass transfer process.

\section{Sample selection}
\label{sec:sample}

We select single-lined (``SB1'') candidates from the \texttt{binary\_masses} table in the {\it Gaia} archive. This table -- for sources near the main sequence only -- contains the results of inferring the masses of both components of a binary through joint fitting of the unresolved source's de-reddened CMD position and the orbital mass function. Details are given in Appendix D of \citet{Arenou2022}. In cases where no consistent solution can be found with two main-sequence stars explaining both a source's CMD position and mass function, a compact object companion is one possible explanation. Other explanations include triple systems, binaries containing undermassive stripped stars that only masquerade as being on the main-sequence \citep[e.g.][]{El-Badry2022_1850}, and spurious orbital solutions.

We select candidates for SB1s containing black holes with the following ADQL query:

\begin{lstlisting}
select * from gaiadr3.binary_masses as bm, gaiadr3.nss_two_body_orbit as nss
where bm.m2_lower > bm.m1_upper 
and bm.m2_lower > 3
and bm.source_id = nss.source_id
and nss.nss_solution_type = 'SB1'
and nss.significance > 20
\end{lstlisting}
Here \texttt{m1\_upper} is the CMD-inferred upper limit of the mass of the RV-variable star with a single-lined binary solution, and \texttt{m2\_lower} is the mass function-inferred lower limit on the mass of the companion, whose contributions were not detected in the {\it Gaia} RVS spectra.

Of the 181,529 sources in the \texttt{nss\_two\_body\_orbit} table with SB1 solutions, 60,271 have mass estimates in the \texttt{binary\_masses} table. Of these, 24,649 pass the cut of \texttt{significance > 20}. For SB1s, \texttt{significance} is defined as the RV semi-amplitude divided by its uncertainty. A larger sample could be studied with a less stringent \texttt{significance} cut, but \citet{Arenou2022} have shown that the increased sample size this affords comes at the expense of much lower purity. 

The query returns 15 sources. One of them, {\it Gaia} DR3 2006840790676091776, is 32 arcsec away from the very bright ($G = 3.86$) star HR 8621. Following \citet{Arenou2022}, we reject this source under the assumption that the RVS spectra were contaminated by the bright companion. The remaining 14 sources are listed in Table~\ref{tab:summary}. As noted by~\citet{Arenou2022}, three objects -- sources 548272473920331136, 1850548988047789696, and 6000420920026118656 -- are eclipsing, and thus unlikely to host compact object companions. However, we retain these sources in our sample for completeness. Except for the eclipsing sources, the resulting sample is identical to the compact object candidate sample selected in Table 10 of \citet{Arenou2022}.

\begin{table*}
\begin{tabular}{llllllllllll}
Source ID  & $P_{\rm orb}$ & $f_m$ & $T_{\rm eff,\,donor}$ & $R_{\rm donor}$ & $T_{\rm eff,\,accretor}$ & $R_{\rm accretor}$ & ${\rm amp}_g$ & $f_{\rm donor}$  & $i$ & $M_{\rm donor}$ & $M_{\rm accretor}$  \\
 & [$\rm days$] & [$M_{\odot}$] & [$\rm kK$] & [$R_{\odot}$] & [$\rm kK$]  & [$R_{\odot}$] & [\%] &  & [$\rm deg$]   &  [$M_{\odot}$] &  [$M_{\odot}$]  \\

\hline
948585824160038912 & 8.202 & 1.47 &  $4.8 \pm 0.1$ & $4.7 \pm 0.1$ & $10.0 \pm 0.2$ &  $1.7 \pm 0.1$ & 5.6 & $0.99 \pm 0.03$ & $73 \pm 3$ & $0.22 \pm 0.03$ & $2.0 \pm 0.1$  \\
2966694650501747328 & 10.398 & 1.16 &  $4.7 \pm 0.1$ & $4.7 \pm 0.1$ & $9.4 \pm 0.2$ &  $2.1 \pm 0.1$ & 2.8 & $0.84 \pm 0.03$ & $60 \pm 3$ & $0.23 \pm 0.03$ & $2.0 \pm 0.1$  \\
2219809419798508544 & 10.865 & 1.29 &  $4.5 \pm 0.1$ & $5.8 \pm 0.1$ & $9.3 \pm 0.2$ &  $2.1 \pm 0.1$ & 5.7 & $0.99 \pm 0.03$ & $68 \pm 3$ & $0.24 \pm 0.03$ & $2.0 \pm 0.1$  \\
5536105058044762240 & 12.177 & 1.18 &  $4.9 \pm 0.2$ & $5.0 \pm 0.2$ & $8.4 \pm 0.3$ &  $2.5 \pm 0.2$ & 2.2 & $0.80 \pm 0.03$ & $66 \pm 3$ & $0.23 \pm 0.04$ & $1.9 \pm 0.2$  \\
5694373091078326784 & 12.876 & 1.66 &  $4.6 \pm 0.2$ & $5.5 \pm 0.2$ & $9.9 \pm 0.3$ &  $1.9 \pm 0.2$ & 2.7 & $0.84 \pm 0.03$ & $75 \pm 3$ & $0.24 \pm 0.04$ & $2.1 \pm 0.2$  \\
548272473920331136 & 13.379 & 1.61 &  $4.7 \pm 0.1$ & $5.3 \pm 0.1$ & $8.8 \pm 0.2$ &  $2.3 \pm 0.1$ & 5.6 & $0.88 \pm 0.01$ & $81 \pm 1$ & $0.18 \pm 0.02$ & $1.9 \pm 0.1$  \\
6734611563148165632 & 14.344 & 1.70 &  $4.8 \pm 0.1$ & $5.3 \pm 0.1$ & $9.7 \pm 0.2$ &  $2.3 \pm 0.1$ & 2.6 & $0.82 \pm 0.03$ & $75 \pm 3$ & $0.18 \pm 0.02$ & $2.2 \pm 0.1$  \\
2933630927108779776 & 14.715 & 1.23 &  $4.4 \pm 0.2$ & $7.2 \pm 0.2$ & $9.7 \pm 0.3$ &  $2.3 \pm 0.2$ & 4.5 & $0.96 \pm 0.03$ & $61 \pm 3$ & $0.27 \pm 0.04$ & $2.2 \pm 0.2$  \\
5243109471519822720 & 14.914 & 1.56 &  $4.6 \pm 0.2$ & $6.6 \pm 0.2$ & $9.4 \pm 0.3$ &  $1.9 \pm 0.2$ & 4.4 & $0.87 \pm 0.03$ & $74 \pm 3$ & $0.28 \pm 0.04$ & $2.0 \pm 0.2$  \\
1850548988047789696 & 15.302 & 1.83 &  $4.6 \pm 0.1$ & $6.0 \pm 0.1$ & $9.6 \pm 0.2$ &  $2.8 \pm 0.1$ & 2.6 & $0.83 \pm 0.01$ & $79 \pm 1$ & $0.23 \pm 0.03$ & $2.3 \pm 0.1$  \\
6000420920026118656 & 15.312 & 2.05 &  $4.5 \pm 0.2$ & $6.2 \pm 0.2$ & $9.5 \pm 0.3$ &  $2.5 \pm 0.2$ & 4.1 & $0.91 \pm 0.01$ & $86 \pm 1$ & $0.19 \pm 0.03$ & $2.2 \pm 0.2$  \\
2197954362764248192 & 17.518 & 1.47 &  $4.3 \pm 0.2$ & $9.0 \pm 0.3$ & $9.7 \pm 0.3$ &  $2.3 \pm 0.2$ & 3.5 & $0.90 \pm 0.04$ & $72 \pm 4$ & $0.46 \pm 0.07$ & $2.2 \pm 0.2$  \\
4514813786980451840 & 22.082 & 1.56 &  $4.3 \pm 0.1$ & $10.5 \pm 0.2$ & $10.0 \pm 0.2$ &  $2.3 \pm 0.1$ & 4.2 & $0.91 \pm 0.04$ & $73 \pm 4$ & $0.44 \pm 0.05$ & $2.2 \pm 0.1$  \\
448452383082046208 & 23.436 & 1.47 &  $4.5 \pm 0.2$ & $9.3 \pm 0.3$ & $10.5 \pm 0.3$ &  $2.4 \pm 0.2$ & 3.1 & $0.87 \pm 0.03$ & $67 \pm 3$ & $0.31 \pm 0.04$ & $2.4 \pm 0.2$  \\
\hline
\end{tabular}
\caption{\label{tab:summary} SB1 sources in {\it Gaia} DR3 with reported \texttt{m2\_lower} > 3\,$M_{\odot}$, \texttt{m2\_lower} > \texttt{m1\_upper}, and \texttt{significance} > 20. $f_m$ is the RV mass function; it sets a lower limit on the mass of the companion to the star we call the donor. $i$ is the inclination,  ${\rm amp}_g$ is the $g-$ band min-to-max ellipsoidal variability amplitude, and $f_{\rm donor} = R_{\rm donor}/R_{\rm Roche\,lobe,\,donor}$ is the donor's Roche lobe filling factor. The {\it Gaia} RVS orbital solution corresponds to the donor. Theses sources are the same as those reported in Table 10 of \citet{Arenou2022}, who fit them as containing a single luminous star. However, all constraints are based on our fitting of the SEDs, light curves, and orbital solution with a binary model, which leads to very different conclusions about the nature of the RV-variable stars. }
\end{table*}

\begin{figure*}
    \centering
    \includegraphics[width=\textwidth]{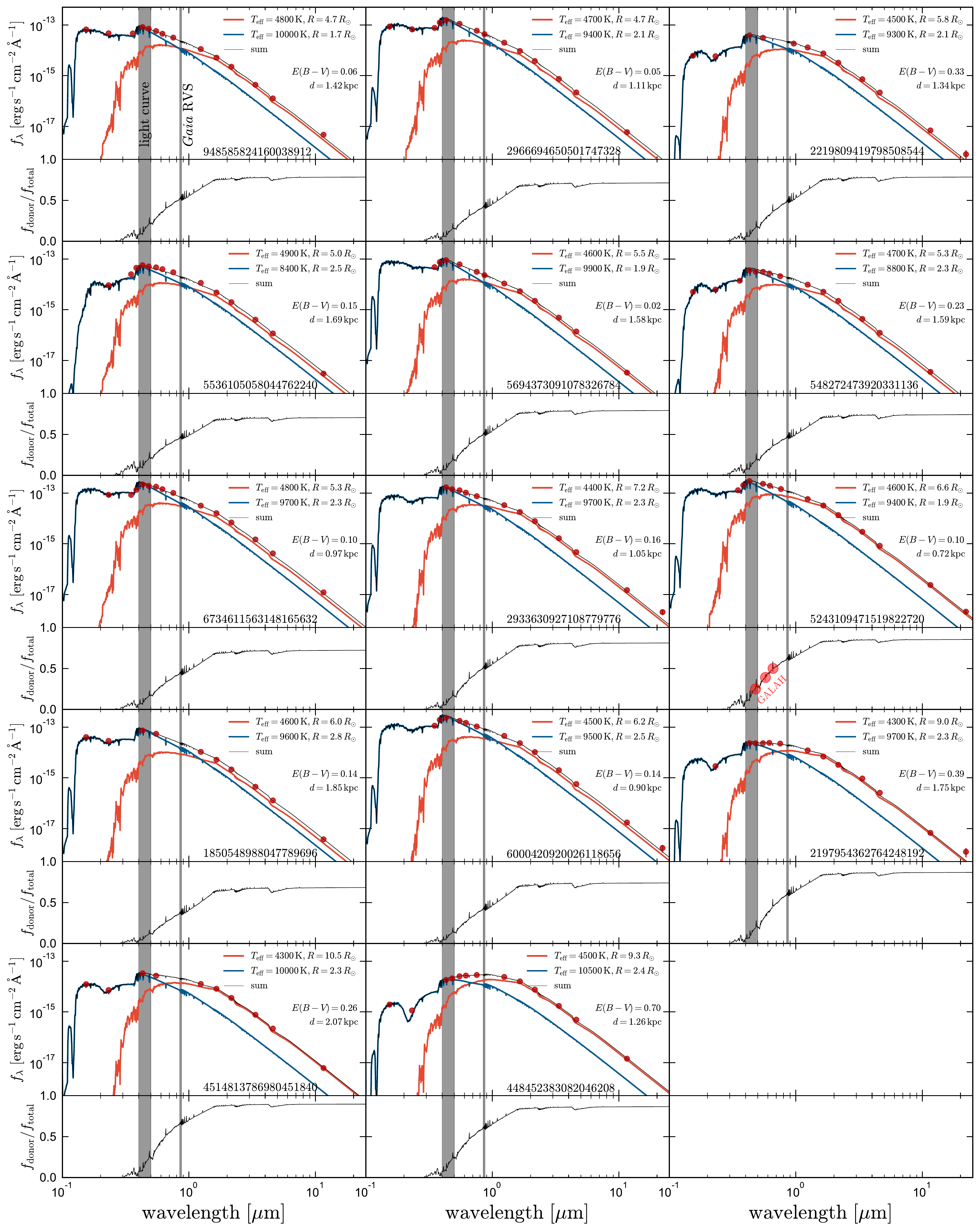}
    \caption{Spectral energy distributions. Red points show observed photometry. Orange and blue lines show models of the accretors (which dominate at short wavelengths) and donors (which dominate at long wavelengths). The {\it Gaia} RV solution corresponds to the donors. Lower sub-panels show the fraction of light contributed by the donors. Shaded regions mark the wavelength of the $g-$ band light curves and {\it Gaia} RVS spectra. The accretors dominate in the $g-$band, diluting ellipsoidal variability. Both components' temperatures and radii are generally well-constrained.  }
    \label{fig:seds}
\end{figure*}
 
\subsection{Spectral energy distributions}
We next retrieved UV-to-IR photometry, as available. All sources were observed by the 2MASS \citep{Skrutskie_2006} and WISE \citep{Wright_2010, Cutri2021} surveys. We supplemented this with $BVgri$ photometry from the APASS survey \citep{Henden2016}, $u$- and $v$-band photometry from SkyMapper \citep{Wolf2018}, and FUV/NUV photometry from GALEX \citep{Martin2005}. The resulting spectral energy distributions (SEDs) are shown in Figure~\ref{fig:seds}, with the reported apparent magnitudes converted to flux densities using the appropriate zeropoints for each survey. 

It quickly became clear that, although these sources fall near the main sequence in the {\it Gaia} CMD, their full SEDs cannot be well-fit by single-star models. The basic issue is that the UV and blue-optical data are well-fit by models that are small ($R \sim 2\,R_{\odot}$) and hot ($T_{\rm eff} \sim 10,000$\,K), but these models under-predict the flux in the near-infrared. This suggested that (at least) two components contribute significantly to the SEDs, so we fit them with a sum of two single-star models.

We predict bandpass-averaged magnitudes using empirically-calibrated solar-metallicity model spectral from the BaSeL library \citep[v2.2;][]{Lejeune1997, Lejeune1998}. We assume $\log g = 4$ for the hotter component and $\log g = 2$ for the cooler component, as motivated by our inference in Section~\ref{light_curves}, but the SEDs are only weakly sensitive to $\log g$ and metallicity. We assume a \citet{Cardelli_1989} extinction law with total-to-selective extinction ratio $R_V =3.1$. We set the reddening $E(B-V)$ based on the \citet{Green2019} 3D dust map for sources with $\delta > -30$ deg and based on the \citet{Lallement2019} 3D dust map for sources farther south. We use \texttt{pystellibs}\footnote{\href{http://mfouesneau.github.io/docs/pystellibs/}{http://mfouesneau.github.io/docs/pystellibs/}} to interpolate between model SEDs, and \texttt{pyphot}\footnote{\href{https://mfouesneau.github.io/docs/pyphot/}{https://mfouesneau.github.io/docs/pyphot/}}  to calculate synthetic photometry. We adopted a prior on the inverse distance based on the zeropoint-corrected parallaxes from {\it Gaia} DR3 (5-parameter solutions, since these orbits are compact and well-described by single-star astrometric models). We then fit each SED using \texttt{emcee} \citep{emcee2013} to sample from the posterior, with the temperatures and radii of both stars, as well as the distance, sampled as free parameters.

The results are shown in Figure~\ref{fig:seds} and listed in Table~\ref{tab:summary}. The parameters of both components are generally well-constrained because the hotter component usually dominates in the UV, while the cooler component dominates in the IR. We show below that these objects are likely products of recent or ongoing mass transfer, so we henceforth refer to the hotter component in each system as the ``accretor'' and the cooler component as the ``donor.''  The fits imply that the donors typically contribute $40-60\%$ of the total light in the {\it Gaia} RVS bandpass, but only 10-30\% of the light in the $g-$ band. 

\subsection{Spectra}
\label{sec:spectra}
We found archival spectra for 4 of the 14 sources. Sources 5243109471519822720 and 6000420920026118656 were observed by the GALAH survey \citep[][]{Buder2021}. The latter source is eclipsing and was observed at phase 0.97, during the ingress of the primary eclipse. Sources 948585824160038912 and 448452383082046208 were observed by LAMOST \citep[][]{Cui2012}.

 \begin{figure*}
    \centering
    \includegraphics[width=\textwidth]{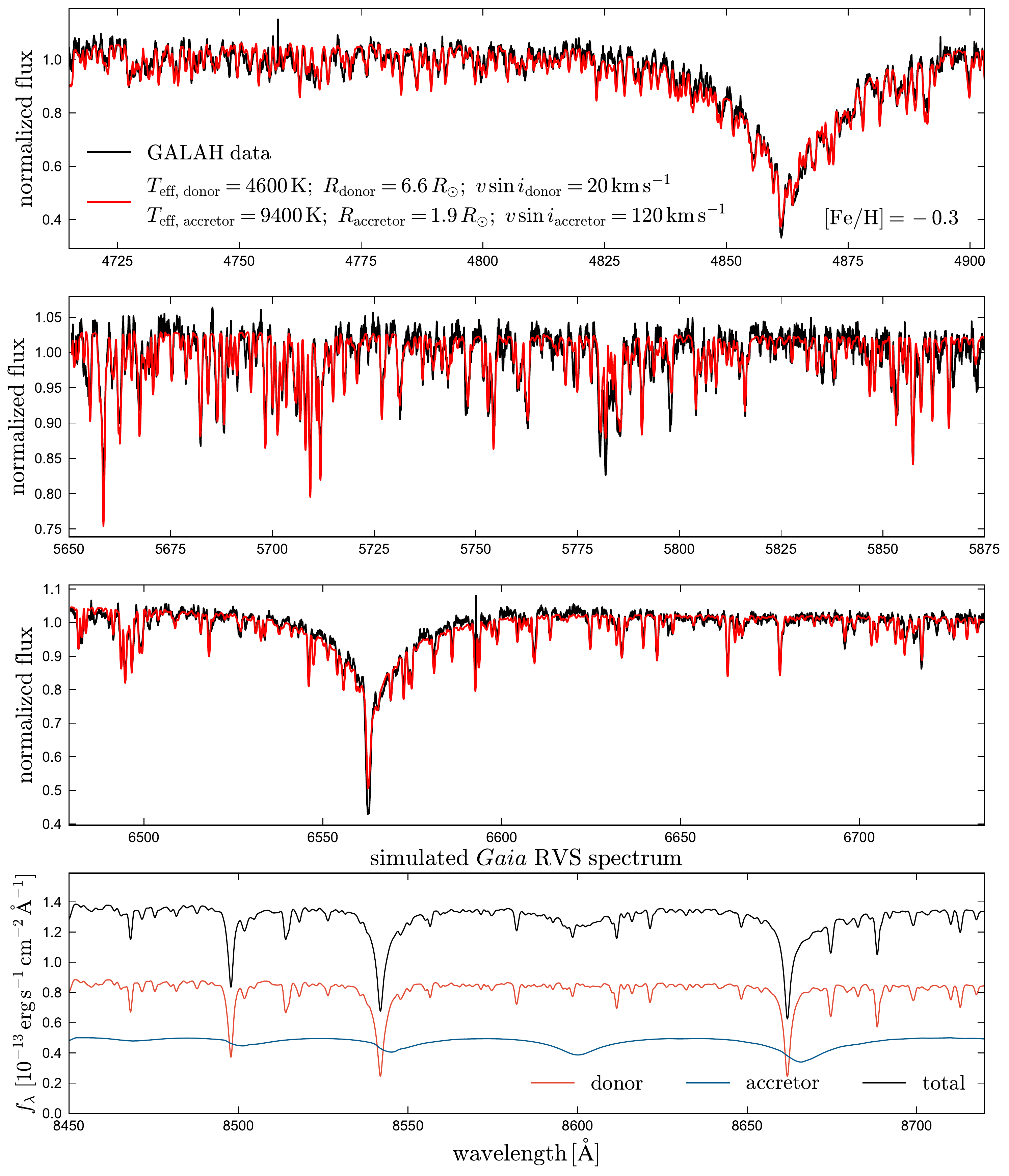}
    \caption{Top three panels show the GALAH survey spectrum of {\it Gaia} DR3 source 5243109471519822720. Red line shows a binary spectral model fit to the data, with the temperatures and radii fixed to the values inferred from the SED. This model provides a good fit to the data, and there is no evidence of emission lines. Bottom panel shows a model spectrum for the same epoch in the {\it Gaia} RVS bandpass. The rapidly-rotating accretor is predicted to contribute $\approx 35\%$ of the light there, but its contributions to the composite spectrum are subtle, making it plausible that the source was not detected as a double-lined binary. }
    \label{fig:galah}
\end{figure*}

We investigated the GALAH spectrum of source 5243109471519822720 (the source which is not in eclipse) in more detail. Figure~\ref{fig:galah} shows cutouts of three wavelength ranges (black) and a binary model spectrum (red) calculated as a sum of two Kurucz models \citep[][]{Kurucz_1970, Kurucz_1979, Kurucz_1993}. Models and data are consistently pseudo-continuum normalized using a first order polynomial fit to each order. We fix the temperatures and radii of both components to the values inferred from the SED, but fit the metallicity, surface gravities, radial velocities, and projected rotation velocities. This yields a good fit to the spectrum, with [Fe/H] = -0.3. The best-fit rotation-velocity of the donor, $v\sin i = 20\,\rm km\,s^{-1}$, is consistent with synchronous rotation and a near-edge-on orbit, since $2\pi R_{\rm donor}/P_{\rm orb}\approx 22\,\rm km\,s^{-1}$. The accretor's rotation velocity, $v\sin i \approx 120\,\rm km\,s^{-1}$, is more rapid, but still slow compared to critical.  We show the light ratio inferred for the three GALAH orders in the appropriate panel of Figure~\ref{fig:seds}; these are consistent with the values predicted by the SED fit. The donor radial velocity is $-92\,\rm km\,s^{-1}$ at JD 2458120.214, in perfect agreement with the {\it Gaia} orbital solution.

In the bottom panel of Figure~\ref{fig:galah}, we show predicted spectra in the {\it Gaia} RVS bandpass. The accretor is predicted to contribute $\approx 35\%$ of the light there, but its contributions are limited to continuum and broad Paschen lines, so it is unsurprising that the source was not detected as an SB2 by the {\it Gaia} pipeline. The mean {\it Gaia} RVS spectrum of this source is published in DR3, and we find that it is in good agreement with the predicted composite spectrum. 

There are no obvious emission lines -- as might be expected for a system with ongoing mass transfer -- in the GALAH spectra. This is not necessarily surprising, because our model suggests  the donor is only $\approx 87\%$ Roche lobe filling (Section~\ref{light_curves}). We discuss the other spectra in Section~\ref{sec:roche_really}. 

\begin{figure*}
    \centering
    \includegraphics[width=\textwidth]{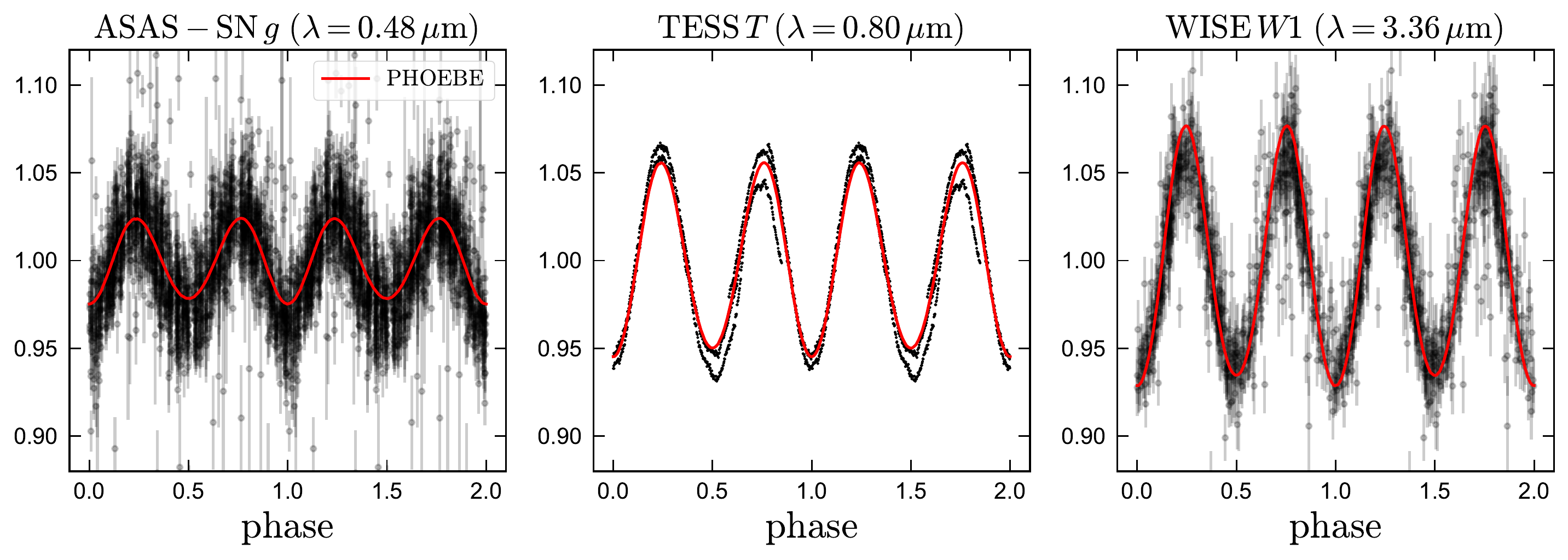}
    \caption{Light curves in three bandpasses for source 2219809419798508544. Left panel shows the blue optical $g-$band light curve, in which the variability amplitude is only a few percent. Center and right panels show light curves at longer wavelengths, where the variability amplitudes are larger. The hot accretor contributes most of the light in the $g-$band, diluting the variability there. Red line shows a PHOEBE light curve model calculated assuming the parameters in Table~\ref{tab:summary}.}
    \label{fig:light_curves}
\end{figure*}

\subsection{Light curves}
\label{light_curves}
We retrieved $g-$ and $V-$band light curves for all sources from the light curve server of the  ASAS-SN survey \citep[][]{Kochanek2017}, which is ideally suited for this sample of bright sources. The data span nearly a decade, with at least several hundred photometric points per object. We also analyzed the $W1-$ and $W2-$band light curves of all targets from the AllWISE and NEOWISE catalogs \citep[][]{Cutri2021, Mainzer2014}; these typically contain a few hundred photometric points spread over more than a decade.

Inspecting Lomb-Scargle periodograms of the light curves, we found a strong peak at half the orbital period reported by {\it Gaia} for all sources in both the ASAS-SN and WISE data. This confirms the {\it Gaia} orbital solutions and suggests that the variability is likely due to ellipsoidal variation. Consistent with this interpretation, 3 sources show eclipses, with the primary eclipse recurring on the orbital period from RVs. Variability in the non-eclipsing systems is close to sinusoidal, with some asymmetry in the depths of adjacent minima due to gravity darkening (e.g. Figure~\ref{fig:light_curves}). 
We fit the non-eclipsing parts of all $g-$ band light curves with a harmonic series and report the peak-to-trough variability amplitudes in Table~\ref{tab:summary}.

The amplitude of ellipsoidal variability depends primarily on the Roche lobe filling factor of the tidally distorted star, which at fixed period depends mainly on the star's mean density \citep[e.g.][]{Eggleton_1983}. Given our constraints on the donor stars' radii, the observed variability thus constrains their masses.

A complicating factor is that the tidally distorted donor stars only contribute a small fraction of the total light in the $g-$band (Figure~\ref{fig:seds}). Their variability is diluted by the smaller and denser accretors, which are not expected to exhibit significant ellipsoidal variability. Fortunately, dilution by the secondary is much less severe in the WISE light curves, where the donors contribute 60-90\% of the light. This is illustrated in Figure~\ref{fig:light_curves}, where we compare the ASAS-SN $g$, {\it TESS} $T$, and WISE $W1$ light curves for a single source. Including the WISE light curves in our fits makes our results much less sensitive to assumptions about the temperatures of both stars.

\begin{figure*}
    \centering
    \includegraphics[width=\textwidth]{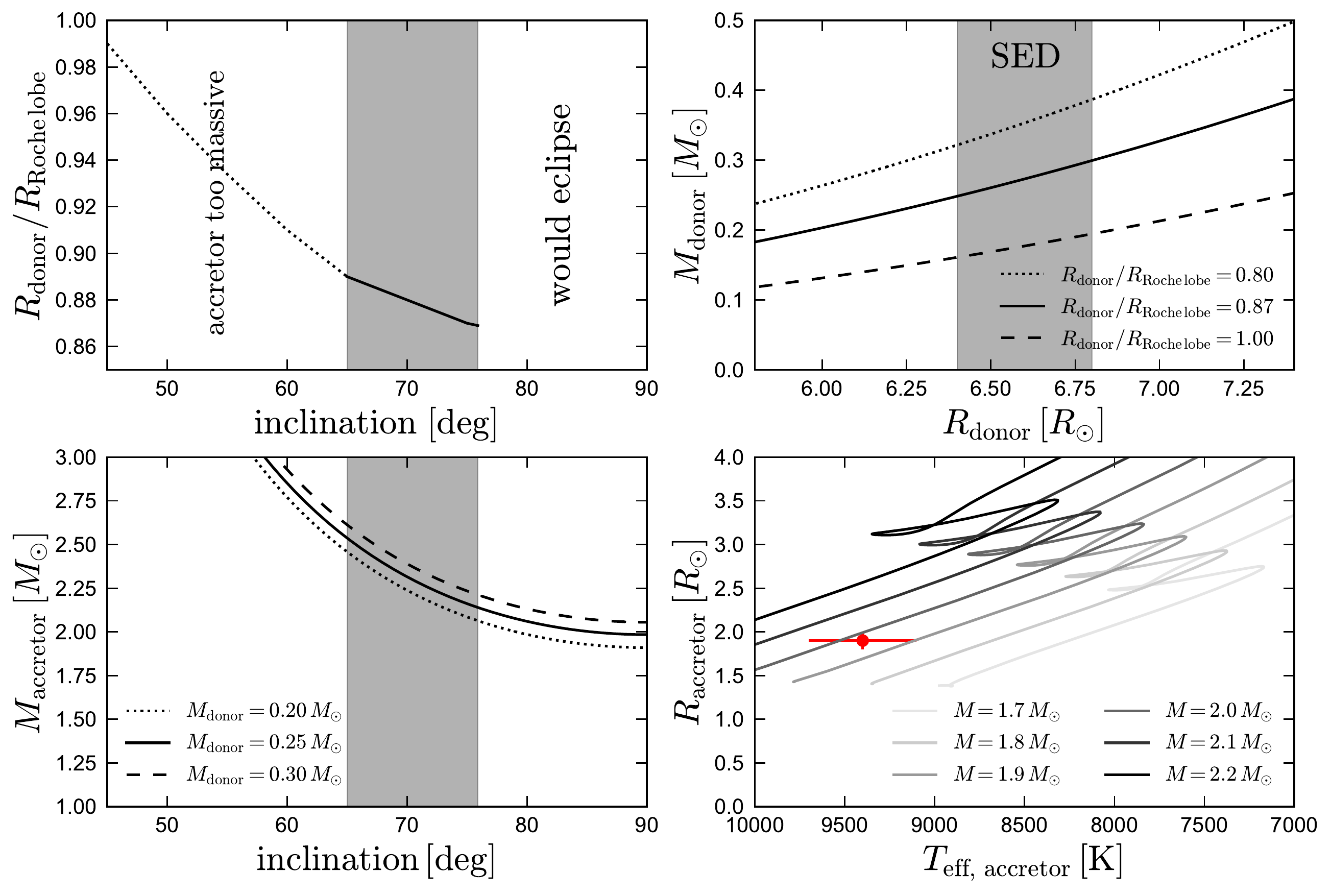}
    \caption{Summary of constraints for one object, {\it Gaia} DR3 source 5243109471519822720. Upper left: donor Roche lobe filling factor vs. inclination. Any point along the solid or dotted line can match the observed ellipsoidal variability amplitude. However, inclinations above 76 deg are ruled out because they would produce eclipses, and inclinations below 65 deg would imply an accretor mass (lower left) inconsistent with the observed accretor parameters (lower right). Upper right: donor mass required to match the observed radius for three choices of Roche lobe filling factor. The light curve constraints imply $M_{\rm donor} \approx 0.28\pm 0.03\,M_{\odot}.$ Lower left: accretor mass implied by the RV mass function for different inclinations. The combination of constraints from the light curve, mass function, and accretor parameters yield tight constraints on the inclination and masses of both components.}
    \label{fig:fourpanels}
\end{figure*}

To account for wavelength-dependent dilution, we calculated model light curves for each system using \texttt{PHOEBE} \citep[][]{Prsa2005}. We set the radii and temperatures of both components to the values inferred from the SEDS, and the mass of the hot component to the value inferred by comparing its temperature and radius to isochrones (e.g. Figure~\ref{fig:fourpanels}, lower left). Gravity- and limb-darkening coefficients are calculated self-consistently using \citet{Castelli2003} atmosphere models. We vary the donor mass (which sets the Roche lobe filling factor at fixed period), and the orbital inclination. 

In the regime of interest, where $M_{\rm accretor} \gg M_{\rm donor}$, the amplitude of ellipsoidal variability at fixed period and light ratio constrains the system to a one-dimensional locus in the plane of inclination and donor Roche lobe filling factor, as illustrated in the upper left panel of Figure~\ref{fig:fourpanels}. Absent other information, it does not independently constrain either parameter. Fortunately, other information is available. High inclinations can be excluded by lack of eclipses (or, for eclipsing systems, the eclipse shape constrains the inclination). Given the observed RV mass functions, low inclinations can also be excluded because they would imply accretor masses inconsistent with the values inferred from their temperature and radii. This is illustrated in the lower two panels of Figure~\ref{fig:fourpanels} and allows us to constrain the inclination and donor Roche lobe filling factor independently. Once the Roche lobe filling factor is constrained, the constraint on the donor's radius constrains its mass (upper right panel of Figure~\ref{fig:fourpanels}).

In practice, we fit for the inclination and masses of both components simultaneously, using the method described in \citet{El-Badry2022hd15124}. The likelihood function compares the predicted \texttt{PHOEBE} light curves to the ASAS-SN $g-$band and WISE $W1-$band data, the temperature and radius of the accretor to MIST single-star isochrones \citep[][]{Choi_2016}, and the predicted orbital mass function of the donor to the observed value. We sample from the posterior using \texttt{emcee} \citep{emcee2013}, and perform standard posterior predictive checks to ensure a satisfactory solution. 

Our constraints are listed in Table~\ref{tab:summary}. The most striking result is that all the donors have very low masses, with the best-fit values ranging from 0.18 to 0.46\,$M_{\odot}$. The inclinations are generally relatively high, as expected for a sample selected in part based on high RV mass function. The best-fit accretor masses range from 1.9 to 2.4\,$M_{\odot}$: early A stars. Although all the donors are tidally distorted, fitting implies that they do not all fill their Roche lobes: a majority have filling factors below 1, with the lowest values near 0.8. 

\subsubsection{Do most of the donor stars really not fill their Roche lobes?}
\label{sec:roche_really}
\begin{figure*}
    \centering
    \includegraphics[width=\textwidth]{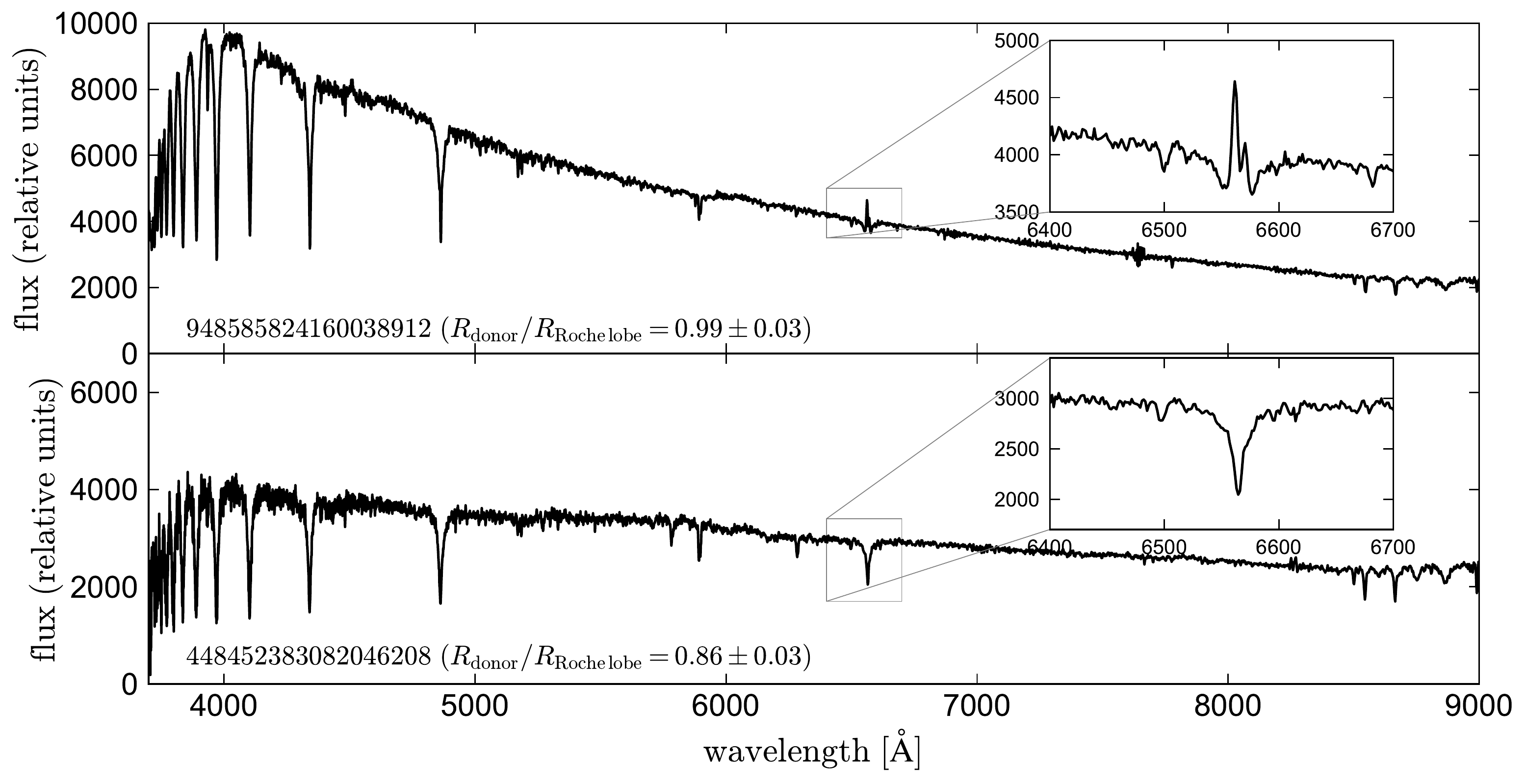}
    \caption{LAMOST spectra of two objects. Top panel shows a source for which our fitting finds that the donor fills its Roche lobe; bottom panel shows one in which we infer that it does not. Double-peaked H$\alpha$ emission is visible only for the Roche-lobe filling source. This likely originates in an accretion disk and is indicative of ongoing or very recently terminated mass transfer. Lack of emission lines in the lower panel suggests very low mass transfer rates in the sources whose donors do not fill their Roche lobes. }
    \label{fig:lamost}
\end{figure*}

It may seem surprising that only three objects in the sample have inferred Roche lobe filling factors consistent with 1, even though they all have $R_{\rm donor}/R_{\rm Roche\,lobe} \gtrsim 0.8$. To investigate this further, we inspected all available archival spectra for evidence of mass transfer. The spectra of the two objects observed by the LAMOST survey are shown in Figure~\ref{fig:lamost}. These show that source 948585824160038912, for which we find $R_{\rm donor}/R_{\rm Roche\,lobe}=0.99\pm 0.03$, has clear double-peaked emission in H$\alpha$. This most likely originates in an accretion disk, consistent with ongoing mass transfer. On the other hand, source 448452383082046208 (for which we found $R_{\rm donor}/R_{\rm Roche\,lobe}=0.87\pm 0.03$)  has no spectral signatures of mass transfer. The same is true for the GALAH spectrum of the eclipsing source 6000420920026118656, for which we found $R_{\rm donor}/R_{\rm Roche\,lobe}=0.86\pm 0.03$.

In summary, the three sources for which we found a donor Roche lobe filling factor inconsistent with 1 have no emission lines, and the source we found to be Roche lobe filling has double-peaked Balmer emission. This supports the hypothesis that mass transfer is halted or slowed in most systems. Follow-up observations of the 10 sources without archival spectra will test this scenario.

\subsubsection{Individual component CMD positions}

 \begin{figure}
    \centering
    \includegraphics[width=\columnwidth]{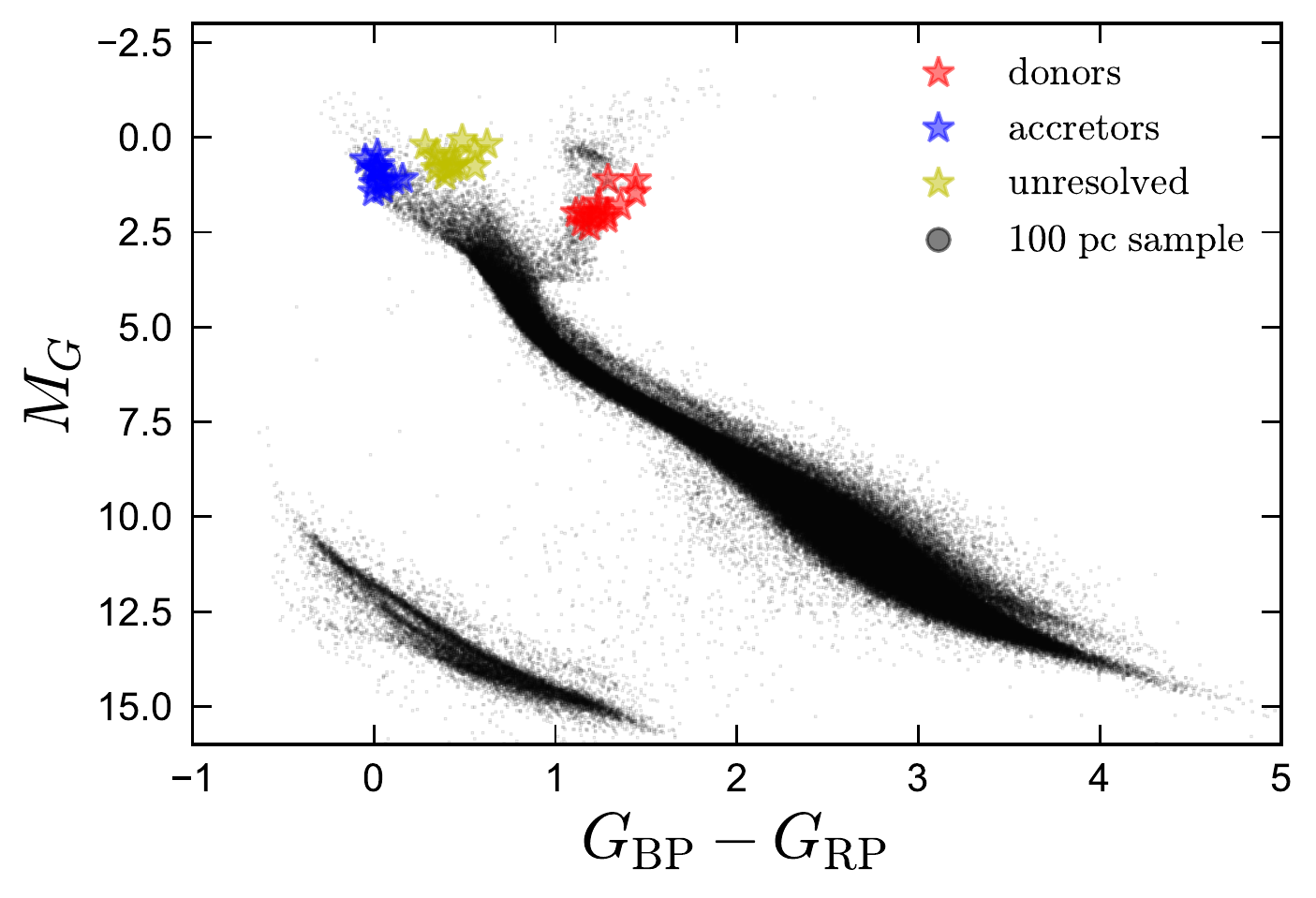}
    \caption{Predicted de-reddened CMD positions of the donors, accretors, and unresolved binaries, as inferred from the SED fits (Figure~\ref{fig:seds}). The rejuvinated accretors all fall near the zero-age main sequence, while the stripped donors are on the lower red giant branch. As unresolved binaries, the sources appear to be slightly evolved main-sequence stars. However, the RV-variable donors have masses of only $0.2-0.3\,M_{\odot}$, ten times lower than inferred when the unresolved sources' CMD positions are interpreted with single-star evolutionary models.   }
    \label{fig:cmd}
\end{figure}

In Figure~\ref{fig:cmd}, we separately plot the inferred {\it Gaia}  CMD positions of both components, and that of the binary when it is observed as an unresolved source. These are calculated from the SED fits in Figure~\ref{fig:seds} and represent de-reddened quantities.  The donors and accretors respectively fall on the lower giant branch and on the unevolved main sequence. We emphasize that no aspect of our fitting required this to occur: the temperatures and radii inferred from the SED fits could have put components on any part of the CMD. When observed together as an unresolved source, the binaries appear in the same region of the CMD as moderately evolved $\sim 2\,M_{\odot}$ single stars. It is striking in Figure~\ref{fig:cmd} how similar all the sources appear to one another.

\subsection{Evolutionary history}
\label{sec:evolhistory}
In all the binaries in our sample, the more evolved component is a low-mass giant, with a typical mass 10 times lower than its unevolved companion. Such a configuration can only be the result of mass transfer, in which the low-mass giant component was initially the more massive star but has since lost most of its mass to the companion. If the companion is able to retain much of the mass in such a scenario, it is expected to be rejuvinated (i.e., moved back toward the main sequence) and perhaps, spun up. 
Translating the presently observed properties of the binaries to their initial conditions is nontrivial, because it depends on poorly understood aspects of mass transfer. In particular, was most of the mass lost by the donor gained by the accretor, or was it ejected from the system? If a significant fraction of the mass was lost, with what angular momentum did it leave the binary? 

Despite these uncertainties, some insight into the systems' evolutionary histories can be gained from toy models. We can first assume mass transfer was fully conservative. In that case, we can work backward from the observed periods and mass ratios using the model from \citet{Soberman1997}. Doing this, we find that the current periods and mass ratios imply initial periods ranging from 0.3 to 3 days, with a median value of about 1 day. 

We next consider non-conservative mass transfer, with some (constant) fraction of the mass lost by the donor being ejected from the vicinity of the accretor. In this case we can again calculate the initial period as a function of mass ratio. For fully non-conservative mass transfer, this yields unphysically small initial periods for most systems, with typical values of 0.1 day and values as small as 0.02 days in the most extreme cases. This scenario can be ruled out because a $\sim 2\,M_{\odot}$ main-sequence star cannot fit inside an orbit with $P_{\rm orb} \lesssim 0.5$ days. This sort of argument rules out mass transfer efficiencies below $\approx 0.5$ for most of  the sample.\footnote{We emphasize that these estimates depend rather sensitively on the current mass ratio, which is as yet only roughly measured. Follow-up modeling of these systems as double-lined binaries would enable a more accurate measurement.}

To explore the mass transfer rates and lifetimes expected in this evolutionary scenario, we calculated a representative binary evolution model using MESA \citep[][]{Paxton_2011, Paxton_2013, Paxton_2015, Paxton_2018, Paxton_2019}. MESA simultaneously solves the 1D stellar structure equations for both stars, while accounting for mass and angular momentum transfer using simplified prescriptions. We assumed an initial composition of $X=0.706, Y=0.28, Z=0.014$ for both stars. We use the \texttt{photosphere} atmosphere table, which uses atmosphere models from \citet{Hauschildt1999} and \citet{Castelli2003}. The \texttt{MESAbinary} module is described in \citet{Paxton_2015}.  Roche lobe radii are computed using the fit of \citet{Eggleton_1983}. Mass transfer rates in Roche lobe overflowing systems are determined following the prescription of \citet{Kolb_1990}. The orbital separation evolves such that the binary's total angular momentum is conserved ($\alpha=\beta=\delta = 0$), as described in \citet{Paxton_2015}.

 \begin{figure*}
    \centering
    \includegraphics[width=\textwidth]{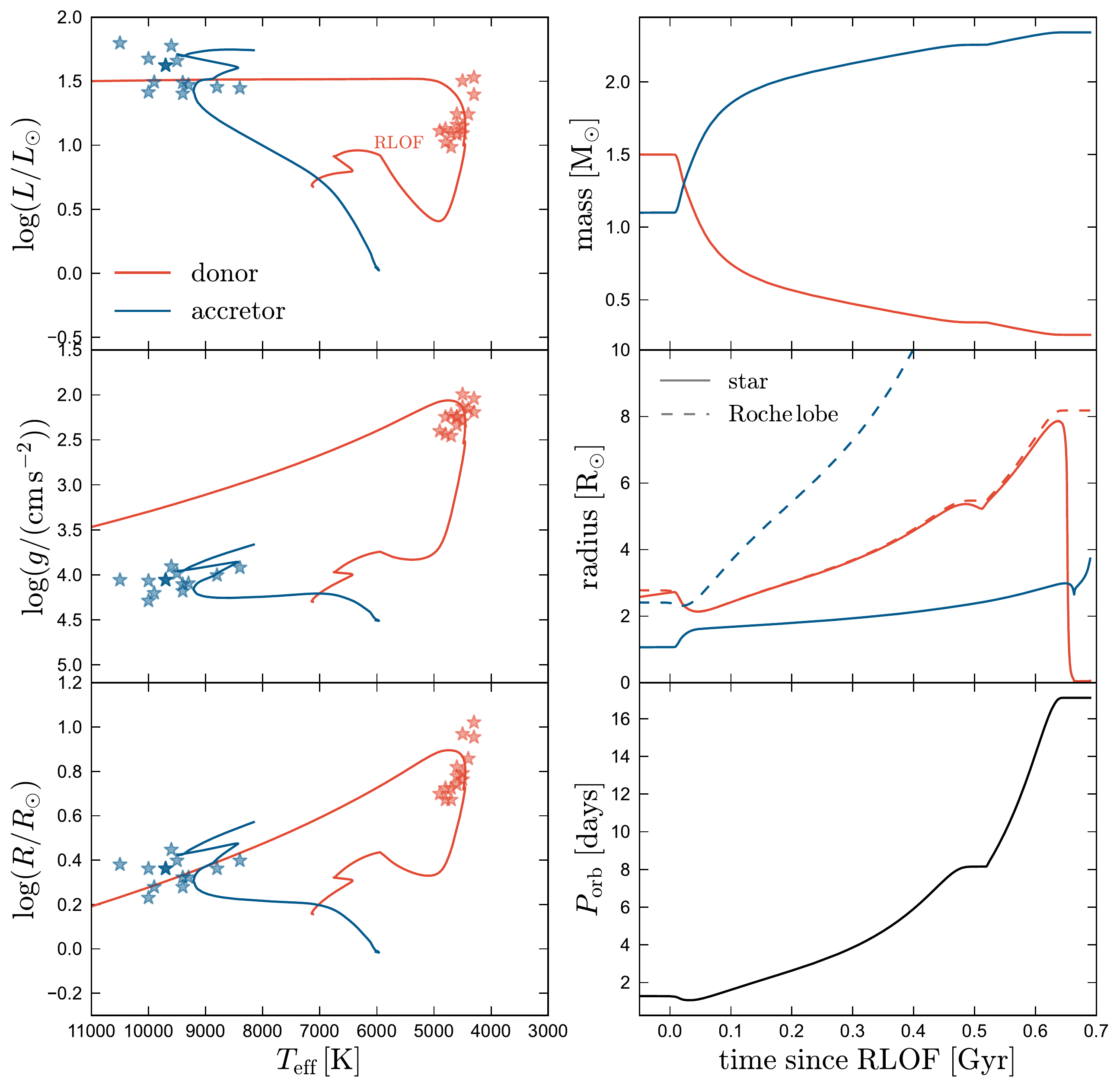}
    \caption{MESA binary evolution model representative of the binaries in the sample. Orange and blue lines show the donor (with inital mass $M_{\rm init} = 1.5\,M_{\odot}$) and accretor ($M_{\rm init} = 1.1\,M_{\odot}$). The initial period is 1.3 days. The donor overflows its Roche lobe (``RLOF'') as a subgiant at an age of 2.3 Gyr. Mass transfer proceeds stably over the next $\approx 0.6$\,Gyr, with the  orbit widening as the donor expands. The mass transfer rate (not shown) peaks at $2\times 10^{-8}\,M_{\odot}\,\rm yr^{-1}$ shortly after initial RLOF and then falls to an average rate of $\approx 10^{-9}\,M_{\odot}\,\rm yr^{-1}$. The donor temporarily detaches from its Roche lobe while it moves up the giant branch, but remains close to Roche lobe filling. Meanwhile, the accretor gains mass and moves up the main sequence. The model resembles the observed binaries (blue and orange stars) near the end of the mass transfer phase, when most of the donor's envelope has been stripped off. The model donor detaches from its Roche lobe and contacts rapidly when only a $\approx 0.2\,M_{\odot}$ helium core with a thin hydrogen envelope remains, leaving behind a helium white dwarf in a 17 day orbit with a $2.3\,M_{\odot}$ main-sequence companion.}
    \label{fig:mesa}
\end{figure*}

We initialize a binary with initial masses $M_1=1.5\,M_{\odot}$ and $M_2= 1.1\,M_{\odot}$ and $P_{\rm orb}=1.3\,\rm days$. The evolution is shown in Figure~\ref{fig:mesa}. The primary overflows its Roche lobe as a subgiant (``case AB'' mass transfer) and subsequently stably transfers its envelope to the accretor over a period of $\approx 0.6$\,Gyr. The mass transfer rate, being set by the donor's nuclear evolution, is never very high: it peaks at $2\times 10^{-8} M_{\odot}\,\rm yr^{-1}$ and is around  $10^{-9} M_{\odot}\,\rm yr^{-1}$ when the donor's properties are similar to those of the observed donors in the {\it Gaia} sample.  

The donor temporarily detaches from its Roche lobe after $\approx 0.5$ Gyr. This occurs when the outward-moving hydrogen-burning shell reaches the hydrogen-rich outer layers of the donor that were not heaving processed by core burning early in the donor's main-sequence evolution, causing the star to shrink temporarily \citep[e.g.][]{Tauris1999}. The duration of this detached episode depends on the initial mass of the donor, the core mass at initial RLOF, and the treatment of mixing. It is relatively short in the model shown in Figure~\ref{fig:mesa}, but is stronger and longer-lasting in some other calculations in the literature \citep[e.g.][]{Podsiadlowski2002}. It seems plausible that the sources for which we find $R_{\rm donor}/R_{\rm Roche\,lobe}\approx 0.9$ are in such a temporary detached phase, but further work is needed to assess the viability of this hypothesis.

Once all but a few $\times\,0.01\,M_{\odot}$ of the donor's envelope has been stripped, it contracts and heats up at constant luminosity, on its way to become a low-mass white dwarf. The model goes through several hydrogen shell flashes (not visible in Figure~\ref{fig:mesa}) during which most of the remaining hydrogen in the envelope is burned. As it accretes the donor's envelope, the model accretor moves up the HR diagram parallel to the main-sequence, also matching the properties of the observed systems. The calculation ends with a low-mass helium white dwarf orbiting an A star: an EL CVn binary \citep[e.g.][]{Maxted2014}. The accretor will soon also overflow its Roche lobe, most likely leading to a common envelope episode, and perhaps, a compact binary containing two helium white dwarfs \citep[e.g.][]{Rappaport2009}.

\section{Summary and discussion}
\label{sec:summary}

We have investigated the nature of single-lined spectroscopic binaries with large mass functions in the upper main sequence of the {\it Gaia} CMD. These are the same sources highlighted in Table 10 of \citet{Arenou2022}. If the {\it Gaia} RVS-detected stars in these sources were normal main-sequence stars, the minimum masses of their companions would exceed $3\,M_{\odot}$ -- larger than the sources' CMD-inferred masses. They thus naturally appear as promising  candidates to host black hole companions.

We find good agreement between the sources' photometric variability and orbital solutions, suggesting that the {\it Gaia} orbital solutions are likely reliable. However, our analysis shows that all the 14 sources we selected are mass-transfer binaries containing highly stripped low-luminosity giants. Their companions are main-sequence A stars, which were likely rejuvinated by accretion. The low-mass stripped stars (``donors'') dominate the light (and in particular, the spectral features) in the far-red {\it Gaia} RVS bandpass, but the companions (``accretors'') dominate in the UV and blue optical (Figures~\ref{fig:seds} and~\ref{fig:galah}). All sources show quasi-sinusoidal light curve variability on half the orbital period, suggestive of ellipsoidal variation (Figure~\ref{fig:light_curves}). The observed amplitude of this variability is low in the optical (few percent in the $g-$band; Table~\ref{tab:summary}), but the true variability amplitudes of the donors are significantly higher, because the companions dilute their light. Joint modeling of light curves and SEDs suggest that some, but not all, of the donors fill their Roche lobes; those that do not have filling factors of $\gtrsim$ 80\%. 

Given the donors' high Roche lobe filling factors, their radii and orbital periods constrain their masses. These have typical values of $0.2-0.4\,M_{\odot}$, much too low to be the result of single-star evolution. We constrain the accretors' masses through comparison of their radii and temperatures to evolutionary models (Figure~\ref{fig:fourpanels}); this suggests that their masses are $1.9-2.4\,M_{\odot}$. Accretor masses inferred thusly are in good agreement with constraints from the RV mass functions, implying that  their masses (unlike those of the donors) can be reasonably inferred with single-star evolutionary models. These companion masses are lower than the values inferred by \citet{Arenou2022}, which were all $M_2 > 3\,M_{\odot}$; this is a result of our lower inferred masses for the SB1-detected star.

The 14 objects in our sample -- which were selected only to have high mass functions and fall near the main-sequence in the CMD -- are all remarkably similar to one another (Figure~\ref{fig:cmd}). Using MESA binary evolution models, we find that they can all be produced by an Algol-like evolutionary scenario with an initial period of order 1 day, an initial primary mass of about $1.5\,M_{\odot}$, and an initial mass ratio $0.6 \lesssim q \lesssim 0.9$ (Figure~\ref{fig:mesa}). Similar scenarios can of course also occur in higher-mass binaries, and in systems with longer initial periods. Indeed, we suspect that they can explain the majority of the spectroscopic binaries with high mass functions identified by {\it Gaia} in which at least one component is a giant.

Several conditions must be satisfied for a mass-transfer binary to end up in the high-mass companion SB1 sample we selected: (a) the donor must contribute a significant fraction of the light in the {\it Gaia} RVS bandpass, (b) the donor must be sufficiently cool that reliable RVs can be measured from the  RVS spectra using templates with $T_{\rm eff} < 8,125\,\rm K$, (c) the unresolved object must fall in a portion of the CMD that overlaps with single-star main-sequence isochrones used in construction of the \texttt{binary\_masses} catalog, and (d) the mass transfer process must be fairly far long, such that $M_{\rm donor}\ll M_{\rm accretor}$ and the mass function is large. Several well-known binaries do indeed meet these criteria and have mass functions above $3\,M_{\odot}$. The most famous is the 3rd magnitude star $\beta$ Lyrae; other members of the class include RX Cas, SX Cas, and V360 Lac (see \citealt{Harmanec2015}, for a recent review of such objects). {\it Gaia} DR3 does not include spectroscopic solutions for any of these systems (or for many massive stars in general; see figure 4 of \citealt{Arenou2022}), but we anticipate that such mass transfer binaries will continue to be the dominant population of high-mass function binaries in future {\it Gaia} data releases. 
 
\citet{Arenou2022} also considered the possibility that these sources might be mass-transfer binaries, but they noted that given the sources' inferred radii, they would fill their Roche lobes only if they had masses below $0.1\,M_{\odot}$, which is implausibly low even in a binary evolution scenario. The resolution to this is that, when fitting the sources' SEDs with binary models, we find larger and cooler donors than implied by single-star fits, leading to higher donor masses. We also find (Section~\ref{sec:roche_really}) that most donors are not quite Roche lobe-filing.

Further investigation of the binaries in this sample is warranted. We found that their present-day periods and mass ratios are unlikely to be the result of highly non-conservative mass transfer, implying that the accretors have gained significant mass. Precise measurements of mass ratios from fitting SB2 models to multi-epoch spectra will tighten constraints on the efficiency of past mass transfer, which is theoretically very uncertain. Given the recent spin-up by accretion, measurement of the accretors' projected rotation velocities will probe the efficiency of {\it spin-down} mechanisms in A-type stars. Searches for line emission associated with accretion will test our finding that a majority of the donors do not fill their Roche lobes. In parallel, spectroscopic observations near ``eclipse'' (even in non-eclipsing systems) can probe for absorption due to the mass transfer stream or the donor's stellar wind \citep[e.g.][]{El-Badry2022unicorn}.

The sources we investigated here turned out not to have compact object companions, but we note that several other avenues remain to select such companions using {\it Gaia} DR3. Most immediately, our search could be extended to sources off the main sequence, and to sources whose SB1 orbital solutions have lower significance. Reducing our \texttt{significance} threshold from 20 to 5 yields an initial sample of 37 rather than 15 candidates. The {\it Gaia} astrometric solutions and ellipsoidal binary light curve fits also contain potentially promising candidates \citep[e.g.][]{Arenou2022, Gomel2022}, which will be followed-up in the coming months and years.
 
\section*{Acknowledgements}
We thank the referee for constructive comments. This project was developed in part at the Gaia F\^ete, held at the Flatiron institute Center for Computational Astrophysics in June 2022.
This work has made use of data from the European Space Agency (ESA) mission Gaia (http://www.cosmos.esa.int/gaia), processed by the Gaia Data Processing and Analysis Consortium (DPAC, http://www.cosmos.esa.int/web/gaia/dpac/consortium). Funding for the DPAC has been provided by national institutions, in particular the institutions participating in the Gaia Multilateral Agreement. HWR acknowledges the European Research Council for the ERC Advance Grant [101054731].

\section*{Data Availability}
Data used in this study are available upon request from the corresponding author. 



\bibliographystyle{mnras}

\bsp	
\label{lastpage}
\end{document}